\documentclass[referee]{cjaa}

\usepackage{graphicx,natbib}
\usepackage{mathrsfs}

\begin{document}
\title{Tail emission from a ring-like jet: its application
to shallow decays of early afterglows and GRB 050709}
\volnopage{Vol.0 (200x)
No.0, 000--000} \setcounter{page}{1}
\author{Yuan-Chuan Zou and Zi-Gao Dai}
\institute{Department of Astronomy, Nanjing University, Nanjing 210093, China.\\
\email{zouyc@nju.edu.cn, dzg@nju.edu.cn}}
\date{Received~~2006 January 30; accepted~~}
%\maketitle

%\begin{abstract}
\abstract{ Similar to the pulsar, the magnetic axis and the spin axis of the
gamma-ray burst source may not lie on the same line. This may cause a ring-like
jet due to collimation of the precessing magnetic axis. We analyze the tail
emission from such a jet, and find that it has a shallow decay phase with
temporal index equal to $-1/2$ if the Lorentz factor of the ejecta is not very
high. This phase is consistent with the shallow decay phase of some early X-ray
afterglow detected by {\it{swift}}. The ring-like jet has a tail cusp with
sharp rising and very sharp decay. This effect can provide an explanation for
the re-brightening and sharp decay of the X-ray afterglow of GRB 050709.
 }
%\end{abstract}
\authorrunning{Y. C. Zou \& Z. G. Dai}
\titlerunning{Tail emission of a ring-like jet}
\maketitle

\section{Introduction}\label{intro}
It is well known that pulsars originate from the core collapse of massive
stars. The average angle between the spinning axis and the magnetic axis of
pulsars is about 27$^\circ$ \citep{leahy91}. Similarly, the spin axis and
magnetic axis of the central engine of the gamma-ray burst may not lie on one
line. As the ejecta may be collimated by the magnetic axis, while the magnetic
axis is processing, so the ejecta may be in a spiral shape at first\citep[][and
reference therein]{fargion05}. As the diversity of the velocities of the ejecta
[as assumed in the standard fireball model of gamma-ray bursts\citep{piran05}],
the spiral ejecta ejected at different times will collide and merge into one
whole shell at last. These collisions just produce internal shocks of gamma-ray
burst. At last, these collisions make the ejecta merge into a ring-shaped jet.
Even if the the ejecta is conical, the baryon-loaded region still be ring-like
\citep{eichler03}. \citet{granot05} and \citet{eichler05} have analyzed
afterglows from ring-like jets. It has also been used to interpret the
$h\nu_{\rm peak}-E_{\rm iso}$ relation\citep{eichler04}.

Tail emission plays an important role at the times when shocks disappear. The
temporal index is $-(2+\beta)$ for a cone-shaped jet, where $\beta$ is the
spectral index of the emission\citep{kumar00, yamazaki05}. Considering the zero
point effect of time, the light curves can be steeper during a short
period\citep{nousek05, zhang05, wu06}.

\citet{nousek05} and \citet{zhang05} have also shown that a shallow decay with
index of about $-1/2$ follows the steep decay for most X-ray afterglows. For
the X-ray afterglow of short burst GRB 050709, there is an unexpected high-flux
point followed by a very steep decay\citep{fox05}. These two observations can
both be explained naturally by considering the tail emission of ring-like jets.
In \S \ref{sec:model} we give the expressions of tail emission from a ring-like
jet. In \S\S \ref{sec:shallow} and \ref{sec:050709}, the shallow decay and
X-ray afterglow of GRB 050709 are analyzed respectively. At last, we summarize
our results in \S \ref{sec:conclusion}.

\section{Model}\label{sec:model}
\begin{figure}
  \includegraphics[width=0.6\textwidth]{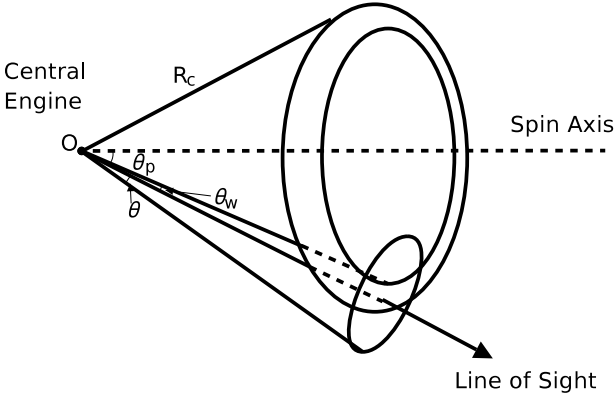}
  \caption{Sketch of a ring-like jet at a distance $R_c$ from the central
  engine, where the main emission is just ceased, while an observer begins
  to receive the tail emission from high latitudes. $\theta_p$ is the half
  opening angle of the ring. $\theta_w$ is the half width angle. $\theta$
  is the latitude of tail emitted region corresponding to the observed time $t$.}
  \label{fig:ring-sketch}
\end{figure}

Considering several ring-like sub-jets emitted from the central engine, they
merge into one whole ring-like jet accompanied with internal shocks. This final
ring with uniform energy density and sharp edges expands with Lorentz factor
$\gamma$, as sketched in Fig. \ref{fig:ring-sketch}. Assuming the radiation
from the ring-like jet begins and ceases at radius $R_c$ (and correspondingly
at time $t_c$) suddenly, we calculate the tail emission from high latitudes of
the ring. The relation is $R_c \simeq 2\eta^2 c t$, where $\eta$ is the mean
Lorentz factor of the internal shocks.

The relation between the latitude angle $\theta$ and the observed time $t$ is
\begin{equation}
  R_c(1-\cos\theta)=c(t-t_c)/(1+z),
\end{equation}
where $z$ is the cosmological redshift. Neglecting the depth of the ejecta 
and the emission from time equal arrival
surface, and defining the emissivity $I'_{\nu'}$ per unit area in the comoving
frame, which is uniform in the whole ring, the flux density in the observer's
frame is
\begin{equation}
  f_{\nu}(t>t_c) =\frac{ I'_{\nu'}}{4\pi D_L^2} \mathscr{D}^2 \frac{{\rm d}S}{{\rm d}t/(1+z)},
  \label{eq:f_nu_general}
\end{equation}
%Doppler因子应该含有红移z
where $D_L$ is the luminosity distance, $\mathscr{D}=1/[\gamma(1-\sqrt{1-1/\gamma^2}
\cos\theta)]$ is the Doppler factor, and ${\rm d}S$ is the emitted area during
a period ${\rm d}t$.

At early times when $\theta < \theta_w$, the tail emission is the same as the
case of an on-axis conical jet, which has been investigated by many authors
\citep{kumar00, fan05}. There are two limiting cases: for $\theta \ll
1/\gamma$,
\begin{equation}
  f_{\nu}(t>t_c) \propto \delta t^0,
  \label{eq:11}
\end{equation}
and for $1 \gg \theta \gg 1/\gamma$,
\begin{equation}
  f_{\nu}(t>t_c) \propto \delta t^{-(2+\beta)},
  \label{eq:12}
\end{equation}
where $\delta t \equiv t-t_c$. Here we consider the emission as a single
power-law profile $I'_{\nu'} \propto \nu'^{-\beta}$, which is valid for the
high frequency emission $\nu' > \max(\nu'_c, \nu'_m)$, where $\nu'_c$ is the
cooling frequency and $\nu'_m$ is the typical frequency of synchrotron
emission.

In the case $\theta > \theta_w$, the width of the ring can be neglected, and
the flux density
\begin{equation}
  f_{\nu}(t>t_c) \propto \mathscr{D}^{-(2+\beta)} \frac{\sin{\theta_p}
  \cos{(\theta/2)}}{\sin{\theta} \sqrt{1-\left(\frac{\sin {(\theta/2)}}
  {\sin {\theta_p}}\right)^2}}.
  \label{eq:f_line}
\end{equation}
There are two limiting cases in which equation (\ref{eq:f_line}) can be
simplified. For $\theta \ll 1/\gamma$,
\begin{equation}
  f_{\nu}(t>t_c) \propto \delta t^{-1/2},
  \label{eq:21}
\end{equation}
and for $1 \gg \theta \gg 1/\gamma$,
\begin{equation}
  f_{\nu}(t>t_c) \propto \delta t^{-(5/2+\beta)}.
  \label{eq:22}
\end{equation}

\section{Shallow Decay of Early X-ray Afterglow}\label{sec:shallow}
Statistics of the early X-ray afterglows has shown that there is a shallow
decay phase with temporal index about $-1/2$\citep{nousek05, zhang05}. This
corresponds to the case: $1/\gamma > \theta > \theta_w$, and can be described
by equation (\ref{eq:21}), where the temporal index is just $-1/2$. As the
general shallow decay lasts from $10^2-10^3$s to $10^3-10^4$s \citep[Fig. 1
in][]{zhang05}, this gives limits: the Lorentz factor of the emitting shell
$\gamma < 7.3 (1+z)^{1/2} R_{c,16}^{1/2} \delta t_{3.5}^{-1/2}$, and the width
of the ring-like jet $\theta_w < 1.4\times 10^{-2} (1+z)^{-1/2} R_{c,16}^{-1/2}
\delta t_{2.5}^{1/2}$. (The conventional donation $Q=Q_k\times 10^k$ is used
throughout this paper.) This implies that the shallow decay component
originates from the shocked shells with low Lorentz factors, while these shocks
may be formed due to ejected sub-shells with different Lorentz factors.

This model can answer the following questions:

Firstly, why is there a steep decay before the shallow decay appears in general
case? In \citet{zhang05}, it is general that the temporal index of this steep
decay is less that $-3$. The answer is that the two power law decays originate
from two different emitting shells with different Lorentz factors. The steep
decay corresponds to the greater Lorentz factor shell, which satisfies
$1/\gamma < \theta$, and the temporal indices are $-(2+\beta)$ or
$-(5/2+\beta)$ corresponding to equations (\ref{eq:12}) and (\ref{eq:22})
respectively. The steep decay may become steeper because of the zero time
selection effect\citep{wu06}.

Secondly, why is there no spectral evolution before and after the break time
from the shallow decay phase to the steep decay phase. This is also mentioned
with spectral index value $\sim -1$ in \citet{zhang05}. It is believed that,
after the break, the afterglow becomes a ``normal'' afterglow. It is possible
that the tail emission phase and the ``normal'' afterglow emission phase are
both in the case $\nu_X > \{\nu_m,\nu_c\}$ (corresponding to the spectral index
$-p/2$) and thus the spectra are the same.

Thirdly, since the shallow phase and the steep phase originate from  different
sources, how to understand the conjunction at the break time \citep[also can be
seen in Fig. 1 in][]{zhang05}? As time goes on, the case converts from
$1/\gamma > \theta > \theta_w$ to $\theta > 1/\gamma > \theta_w$, and then the
light curve of the tail emission decay has a temporal index $-(5/2+\beta)$.
This is steeper than the ``normal'' afterglow with temporal index $\sim -1.2$.
Some time later, the ``normal'' afterglow will exceed the tail emission
definitely, as in the case GRB 050525a \citep[Fig. 1 in][]{nousek05} (at about
3000s, there is a steep decay). However, GRB 050315\citep{vaughan05} can be
classified into the case that ``normal'' afterglow exceed the shallow tail
emission before the tail emission breaks to steep phase.
% while some other breaks without jumps may just be coincident cases.

\section{X-ray Afterglow of GRB 050709}\label{sec:050709}

\begin{figure}
  \includegraphics[width=0.8\textwidth, angle=270]{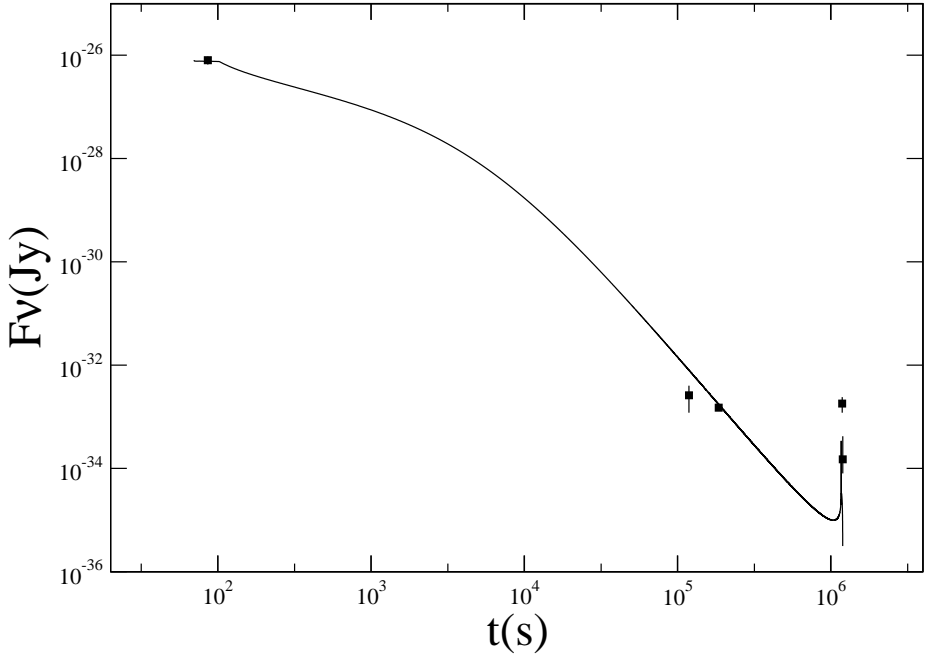}
  \caption{Tail emission fit to the X-ray emission from GRB 050709.
  The five observed data are taken from \citet{fox05}.}
  \label{fig:050709}
\end{figure}

GRB 050709 is a short burst with duration 0.3 s, and five points of X-ray
emission  after the burst were obtained by Swift and Chandra \citep{fox05}.
Figure \ref{fig:050709} shows the fit by assuming that the latter four points
are the tail emissions from the first point, with parameters $R_c=7.7\times
10^{16}$cm, $\gamma=15.5$, $\beta=1.1$, $\theta_p=0.5$ and $\theta_w=0.005$. As
the first X-ray point occurs at time about 100s, the radius $r \simeq 2\eta^2 c
t/(1+z) \simeq 5.2 \times 10^{16} \eta_{2}^2 t_2$cm, is consistent with the
value of the parameter $R_c$. As a short burst has less total energy than a
long burst does, the ejected shell can be decelerated quickly. The Lorentz
factor at $R_c$ is $\gamma \simeq {E_{\rm{iso}}}/({\pi R_c^3 n m_p c^2}) \simeq
26 E_{\rm{iso},50}^{1/2} n_1^{-1/2} R_{c,16.5}^{-3/2}$, where the external
medium density $n$ is chosen equal to 1$\rm{cm}^{-3}$ because the host is a
star-forming galaxy. Therefore, the parameters chosen to fit the X-ray data are
reasonable for this short burst.

We can see four stages for this tail light curve: first, a horizontal  phase
corresponds to the case $\theta < \theta_w < 1/\gamma$; second, a shallow decay
with temporal index $-1/2$ corresponds to the case $\theta_w  < \theta <
1/\gamma$; third, a sharper decay with temporal index $-(2.5+\beta)$
corresponds to the case $\theta_w < 1/\gamma < \theta$; and finally, a tail
cusp with sharp rising and very sharp decay, which comes from the end of the
ring.

We should note that the solid line doesn't fit the data very well, especially
that the tail cusp of the model can't reach to the observed data. This shortage
may be overcome by considering a non-uniform ring-like jet or some other
mechanism. However, its unique feature, which the emission after the tail cusp
decays very sharply, is consistent with the last two observed points. On the
other hand, it is possible that the second, third and fifth points in this
figure belong to the ``normal'' afterglow from an external shock.

\section{Conclusions}\label{sec:conclusion}

Enlightened from pulsars, we suggest that the magnetic axis and the spin axis
of a gamma-ray burst source point to different orientations. The ejecta along
the magnetic axis will form a ring finally. Gamma-ray emission will be observed
if the observer locates in the solid angle of the ring. We have investigated
the tail emission from a ring-like jet. We find that the early shallow decay
phase and the late re-brightening of the X-ray emission of GRB 050709 can be
explained.

Note that the shallow decay phase is only possible in the low Lorentz factor
cases. For the case $1/\gamma < \theta_w$, only the steep one appears. As the
tail emission from the shells with high Lorentz factors decays very quickly,
the main emissions will be dominated by the slower shells at later times.

YCZ thanks helpful discussions with Jia Wang and Xuefeng Wu. This work was
supported by the National Natural Science Foundation of China (grants 10233010
and 10221001).

\end{document}